\documentclass[12pt]{spieman}  
\usepackage{amsmath,amsfonts,amssymb}
\usepackage{graphicx}
\usepackage{setspace}
\usepackage{tocloft}
\usepackage{threeparttable}
\usepackage{lineno}

\title{Multiplexing lobster-eye optics: a concept for wide-field X-ray monitoring}

\author[a,b,c*]{Toru Tamagawa}
\author[b,c]{Keisuke Uchiyama}
\author[d]{Ryota Otsubo}
\author[d]{Tatsuya Yuasa}
\author[b,c]{Yuanhui Zhou}
\author[a,b]{Tatehiro Mihara}
\author[d]{Yuichiro Ezoe}
\author[d]{Masaki Numazawa}
\author[d]{Daiki Ishi}
\author[d]{Aoto Fukushima}
\author[d]{Hikaru Suzuki}
\author[d]{Tomoki Uchino}
\author[d]{Sae Sakuta}
\author[e] {Kumi Ishikawa}
\author[a] {Teruaki Enoto}
\author[f] {Takanori Sakamoto}
\affil[a]{RIKEN Cluster for Pioneering Research, 2-1 Hirosawa, Wako, Saitama 351-0198, Japan}
\affil[b]{RIKEN Nishina Center, 2-1 Hirosawa, Wako, Saitama 351-0198, Japan}
\affil[c]{Tokyo University of Science, 1-3 Kagurazaka, Shinjuku, Tokyo 162-8601, Japan}
\affil[d]{Tokyo Metropolitan University, 1-1 Minami-Osawa, Hachioji, Tokyo 192-0397, Japan.}
\affil[e]{Japan Aerospace Exploration Agency, Institute of Space and Astronautical Science, 3-1-1 Yoshino-dai, Chuo-ku, Sagamihara, Kanagawa 252-5210, Japan}
\affil[f]{Aoyama Gakuin University,
5-10-1 Fuchinobe, Chuoku, Sagamihara, Kanagawa 252-5258, Japan}

\cftpagenumbersoff{figure}
\cftpagenumbersoff{table} 
\begin{document} 
\maketitle

\begin{abstract}
We propose a concept of multiplexing lobster-eye (MuLE) optics to achieve significant reductions in the number of focal plane imagers in lobster-eye (LE) wide-field X-ray monitors. 
In the MuLE configuration, an LE mirror is divided into several segments and the X-rays reflected on each of these segments are focused on a single image sensor in a multiplexed configuration.
If each LE segment assumes a different rotation angle, the azimuthal rotation angle of a cross-like image reconstructed from a point source by the LE optics identifies the specific segment that focuses the X-rays on the imager.
With a focal length of 30~cm and LE segments with areas of 10 $\times$ 10 cm$^2$, $\sim$1~sr of the sky can be covered with 36 LE segments and only four imagers (with total areas of 10 $\times$ 10 cm$^2$). 
A ray tracing simulation was performed to evaluate the nine-segment MuLE configuration.
The simulation showed that the flux (0.5 to 2~keV) associated with the 5$\sigma$ detection limit was $\sim$2 $\times$ 10$^{-10}$ erg cm$^{-2}$ s$^{-1}$ (10 mCrab) for a transient with a duration of 100~s.
The simulation also showed that the direction of the transient for flux in the range of 14 to 17~mCrab at 0.6~keV was determined correctly with 99.7\% confidence limit. 
We conclude that the MuLE configuration can become an effective on-board device for small satellites for future X-ray wide-field transient monitoring.
\end{abstract}


{\noindent \footnotesize\textbf{*}Toru Tamagawa,  \linkable{tamagawa@riken.jp}}

\begin{spacing}{1}

\section{Introduction}
\label{sect:intro}

Wide-field X-ray monitors have been proven to be indispensable devices in time-domain astronomy in recent years.
The precise and immediate localization of transient phenomena is critical for the revelation of their origin.
For example, quick localization of gamma-ray bursts (GRBs) revealed the origin of GRBs with long durations as collapsers\cite{Vanderspek:2004cr}.
Gravitational waves from a neutron star merger were detected\cite{Abbott:2017kt} in 2017 and demarcated the onset of multimessenger astronomy\cite{Abbott:2017it}.
In 2018, follow-up observations were carried out for the neutrino burst detected by IceCube, and the origin of this event was localized to an active galactic nucleus\cite{Anonymous:2018ve}. 
In 2021, the large synoptic survey telescope\cite{Ivezic:2019ch} will begin its observations and will generate several million alerts per night\cite{Ridgway:2014kb}.
Identifying the high-energy counterparts of these visible transients is important for elucidating their origins.
Correspondingly, in multimessenger astronomy, the use of a device that constantly monitors the universe with a wide field-of-view (FoV) in the X-ray energy band is essential.

Coded masks are used for wide FoV missions such as \textit{INTEGRAL}\cite{Winkler:2003er}, \textit{Swift/BAT}\cite{Krimm:2013ch}, \textit{HETE/WXM}\cite{Shirasaki:2003wu}, and \textit{BeppoSAX/WFC}\cite{Jager:1997gk}, but in principle it is difficult to increase the sensitivity because of the interference caused by the diffuse cosmic X-ray background (CXB).
The all-sky monitors on-board \textit{RXTE}\cite{Levine:1996cy} and \textit{MAXI} \cite{Matsuoka:2009dk} improved the detection sensitivities by narrowing the FoV with pinhole camera or slit techniques, and yielded excellent performance in the observation of faint X-ray sources.
To compensate for the improved sensitivity, the sky coverage of a moment was restricted to a few \% of the entire sky.

Lobster-eye (LE) optics \cite{Angel:1979tg} represents the best possible observation equipment for missions that require a wide FoV and increased sensitivity.
The LE optics reduces the influence of the CXB by focusing, and concurrently securing a broad FoV.
Several X-ray astronomical satellite missions, such as \textit{Einstein Probe} \cite{Yuan:2018ec}, \textit{ISS-Lobster}\cite{Camp:2013ks}, and
\textit{HiZ-GUNDAM}\cite{10.1117/12.2055041}
employ the LE optics.
A disadvantage of the LE optic is the necessity for large-sized imagers at the focal plane.
For example, the all-sky monitor mission \textit{LOBSTER} requires a detector area spanning 5000~cm$^2$ to cover $\sim$1/4 of the entire sky \cite{Priedhorsky:1996kv}.
In this study, we describe the design, feasibility, and performance evaluation of a newly proposed idea of LE optics to reduce the number of imagers.

\section{Concept of reduction of focal plane imagers}

\begin{figure}
\begin{center}
\begin{tabular}{c}
\includegraphics[width=12cm]{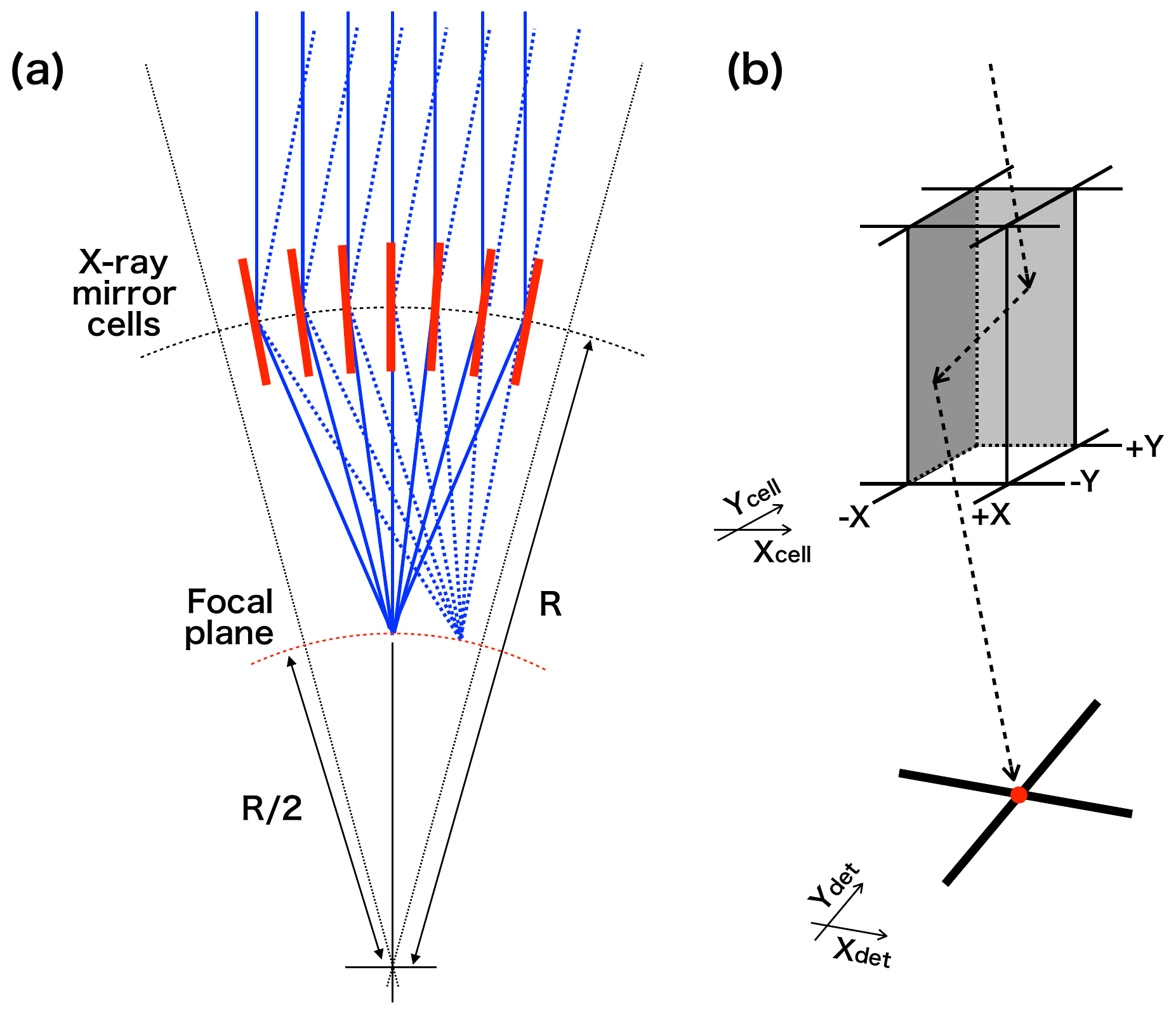}
\end{tabular}
\end{center}
\caption
{\label{fig:lobster_eye}
(a) Schematics of the LE optics showing the X-ray mirror cells mounted on a curved spherical surface with a radius $R$ and focal plane detectors at a focal length of $R$/2.
X-rays from different positions in the sky are focused on different locations on the focal plane.
(b) A square, hollow cell and the path of an X-ray scattered on different planes +Y and -X.
The X-ray photon is focused on the center of a cross-like image generated by the LE optics on a focal plane detector.
The photons reflected on the mirror once are focused on the cross-like arm foci.}
\end{figure}

The LE mirror consists of many square, hollow cells that operate as X-ray reflectors tiled on a curved sphere with a radius \textit{R}, as shown in Fig.~\ref{fig:lobster_eye}.
X-rays that originate from a point source are reflected twice on the adjacent walls of a square hollow cell (Fig.~\ref{fig:lobster_eye}b) and are focused on a point on the focal plane with a radius of \textit{R}/2.
When the incident angle of the X-rays is different, the X-rays are focused on another location on the focal plane.
In combination with the image sensors placed at the focal plane, the LE optics realizes X-ray imaging with wide FoV that is not achievable with any other standard X-ray mirror optics.
As shown in Fig.~\ref{fig:lobster_eye}b, since the X-rays reflected only once in the X (Y) surface of a cell are focused in the X$_\mathrm{det}$ (Y$_\mathrm{det}$) direction but not in the Y$_\mathrm{det}$ (X$_\mathrm{det}$) direction, the focus should be a line along Y$_\mathrm{det}$ (X$_\mathrm{det}$). Thus, those photons are focused on cross-like arm foci.
To cover the entire FoV of the LE mirror, which is the opening angle of an LE mirror segment as described in Appendix~\ref{app:lobster_eye_optics}, large image sensors covering a 1/4 size of the area of the LE mirror are required at the focal plane.
However, imagers with large areas are sometimes unsuitable for a small satellite mission because they consume non-negligible satellite resources, such as electrical power, computer power, and data downlink bandwidth, and may cause cooling problems. 

\begin{figure}
\begin{center}
\begin{tabular}{c}
\includegraphics[width=8cm]{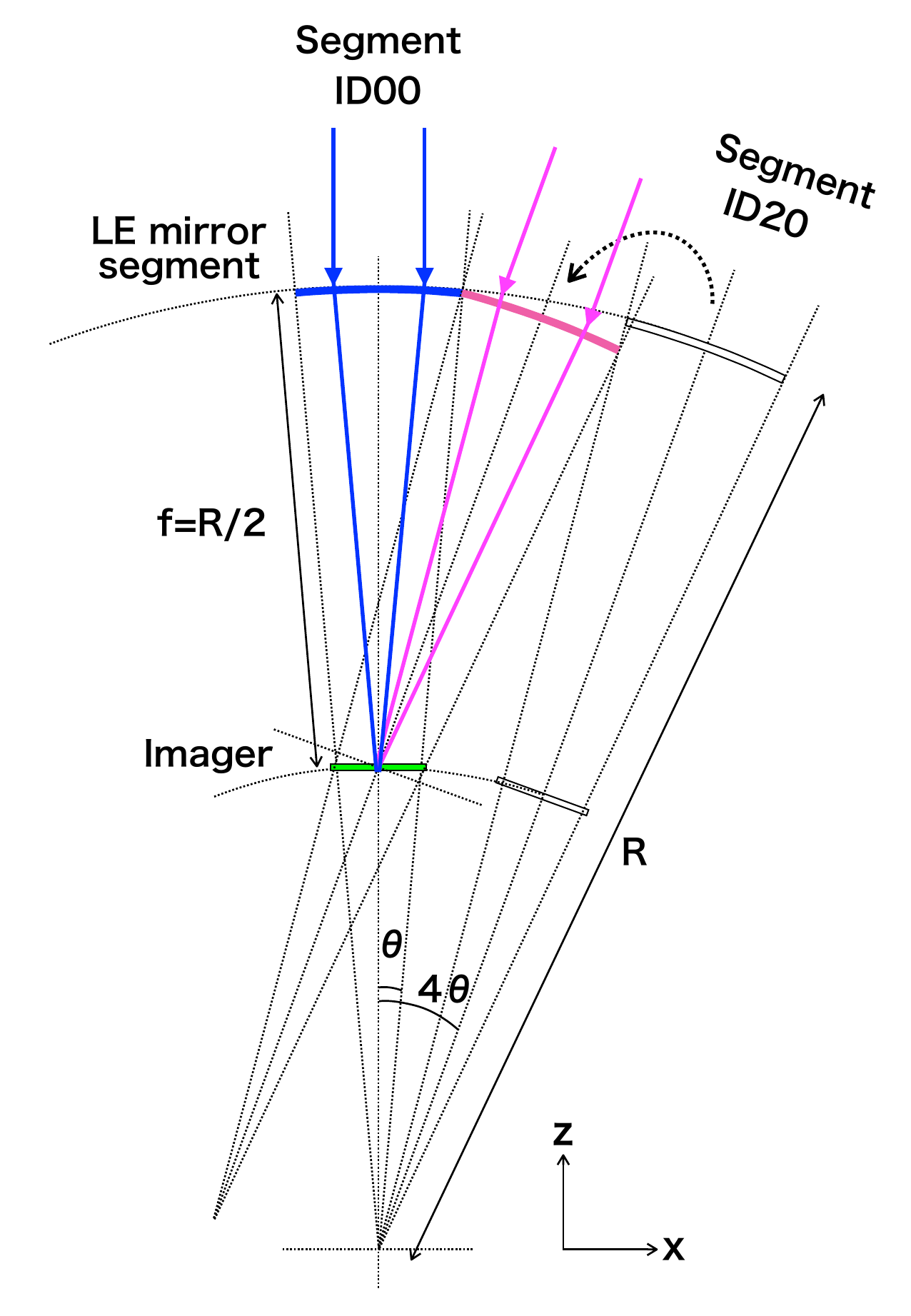}
\end{tabular}
\end{center}
\caption{\label{fig:leoptics}
Conceptual design of the MuLE optics to achieve reductions in the number of imagers.
A single imager is shared between ID00 and ID20 LE mirror segments.}
\end{figure}

To overcome these disadvantages, we propose a new configuration in which the LE mirror is divided in several segments and the X-rays reflected on each segment of the mirror are focused on a single, small image sensor, as shown in Fig.~\ref{fig:leoptics}.
If we define the opening angle of an LE segment as 2$\theta$, the LE segment ID20 in Fig.~\ref{fig:leoptics}, which is 4$\theta$ away from the central segment (ID00), can be moved right next to the ID00 segment.
We refer to this configuration in this study as "multiplexing lobster-eye (MuLE)" optics.
To specify a LE segment, we use the notation ID$n_\mathrm{x}n_\mathrm{y}$, where $n_\mathrm{x}$ and $n_\mathrm{y}$ are indices used to represent the distance of the segment $2n\theta$ away from ID00 in the x and y directions, respectively.
Negative integers are represented with a bar. 
For example, $\bar{n}$ implies $-n$.
The similar configuration was adopted by the ABRIXAS mission \cite{Trumper:1998ip} in which one CCD camera was shared by seven X-ray mirrors.
Their design was to drop different FoVs to different areas of the imager, but in our concept different FoVs are dropped to the same area of an imager.

How can we distinguish two stellar objects focused by different LE segments on one imager?
As shown in Fig.~\ref{fig:lobster_eye}b, a point source focused by an LE segment shows a cross-like response on the imager.
The azimuthal rotation angle of the cross-like arm foci on the imager is exactly the same as that of the square hollow cells of the LE segment around the central optical axis of the segment.
By giving each segment a different azimuthal rotation angle, point sources focused by different mirror segments form cross-like arm foci with different azimuthal rotation angles.

\begin{figure}
\begin{center}
\begin{tabular}{c}
\includegraphics[width=15cm]{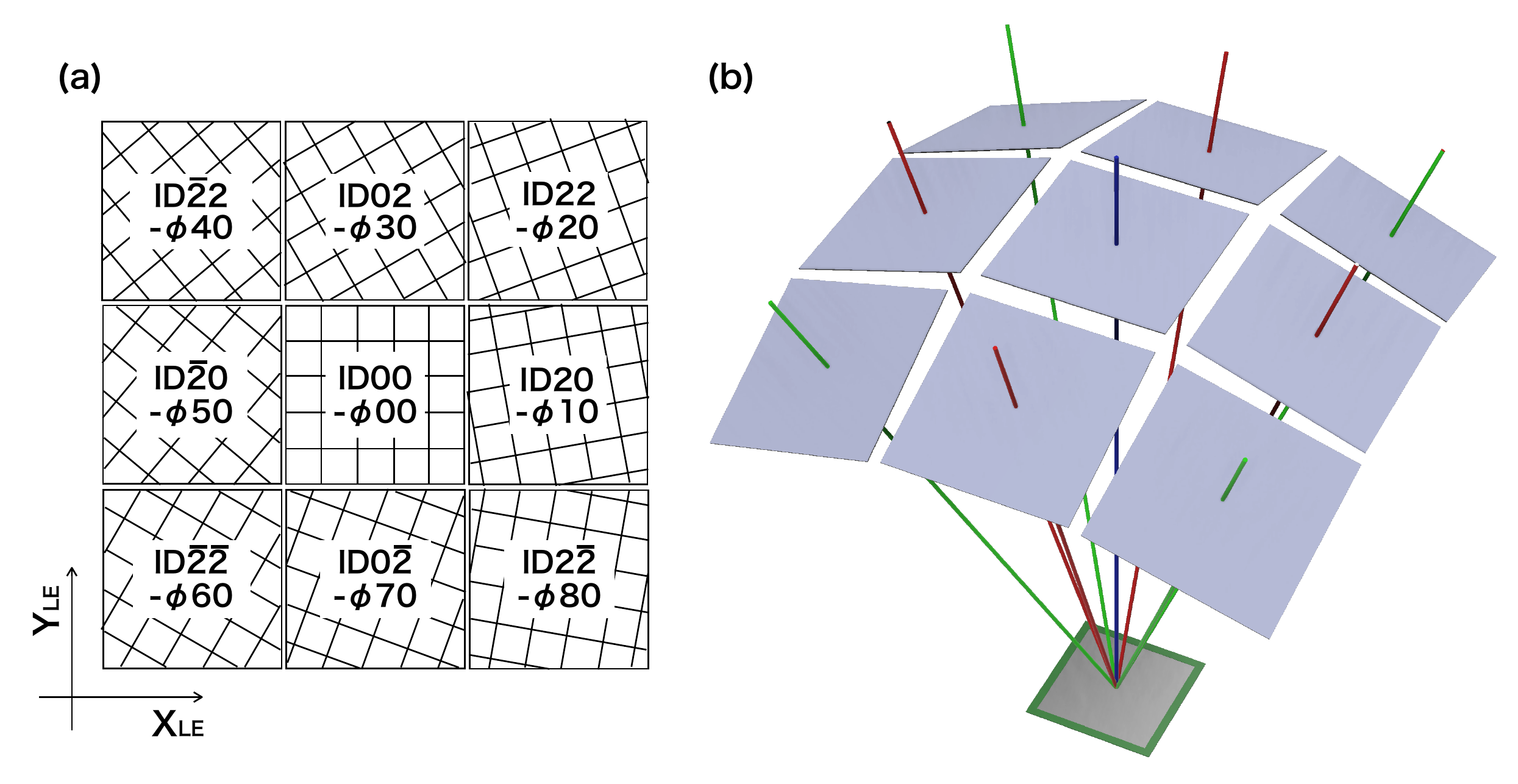}
\end{tabular}
\end{center}
\caption{ \label{fig:le_module_config}
(a) The configuration of nine-segment MuLE optics.
The azimuthal rotation angle of the square, hollow cells of each segment is shifted by 10$^{\rm \circ}$.
The numbers begin with $\phi$ after the use of the LE segment ID to represent the azimuthal rotation angles around the optical axes of the LE segments. 
The cell size of each segment is exaggerated.
(b) Three-dimensional modeling of the nine-segment MuLE optics.}
\end{figure}

We estimated the total FoV covered by the MuLE optics.
The half angle $\theta$ of the FoV of each LE segment was defined as $\theta=\sin^{-1}(L/2R)$, where each LE segment had an area of $L\times L$.
If we consider specific values $R=60$~cm and $L=10$~cm, the FoV of each segment becomes $9.6^{\bf\circ}$ $\times$ $9.6^{\bf\circ}$.
One of possible configurations of the MuLE optics consists of nine tiled segments, as shown in Fig.~\ref{fig:le_module_config}, in which an azimuthal rotation angle of each LE segment increments 10$^{\rm \circ}$ from 0$^{\rm \circ}$ to 80$^{\rm \circ}$.
The numbers $\phi$00--$\phi$80 in Fig.~\ref{fig:le_module_config}a indicate the azimuthal rotation angles of the LE segment cells around the optical axis.
It is not difficult to manufacture such mirrors with current technology.

\begin{figure}
\begin{center}
\begin{tabular}{c}
\includegraphics[width=10cm]{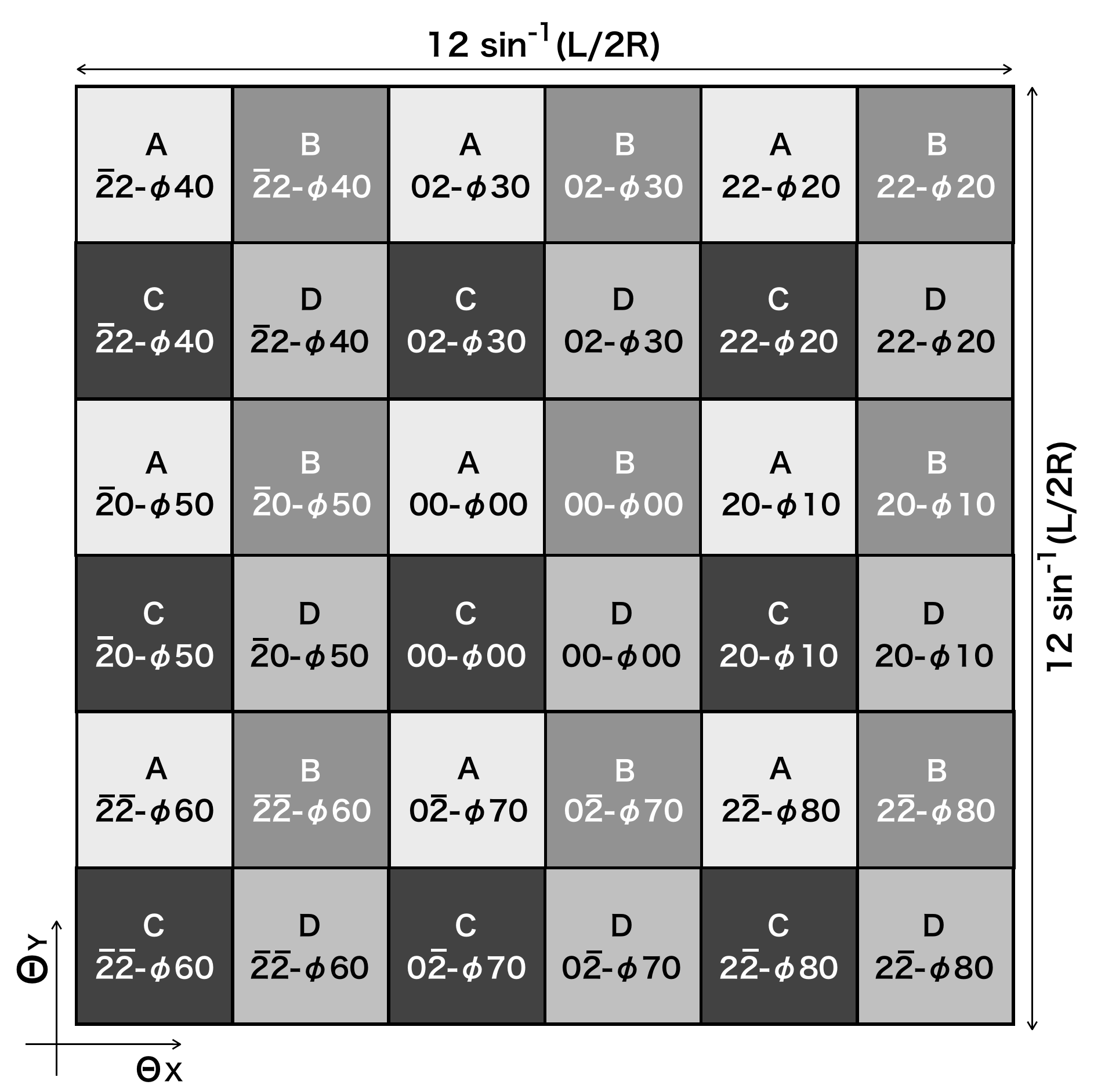}
\end{tabular}
\end{center}
\caption{\label{fig:fieldofview}
FoV covered with four units of the nine-segment MuLE optics.
A to D denote the unit numbers.
For $R=60$~cm and $L=10$~cm, the angular span of 57.4$^{\rm \circ}$ $\times$ 57.4$^{\rm \circ}$ ($\sim$1~sr) of the sky is covered with 36 LE segments and four image sensors.
}
\end{figure}

As observed from Fig.~\ref{fig:leoptics}, the FoV covered by ID20 is not the continuous tiling of ID00 but a tiling configuration at every other position in the sky coordinate system.
When four units of the nine-segment MuLE are used, a sky area of 57.4$^{\bf\circ}$ $\times$ 57.4$^{\bf\circ}$ can be covered, as shown in Fig.~\ref{fig:fieldofview}.
Each unit is named A to D, and an imager is installed directly under ID00 of each unit.
The four ID00s of units A to D are installed offset from each other by 2$\theta$ in the x and y directions.
Since each segment of the nine-segment MuLE covers the FoV every 4$\theta$ (Fig.~\ref{fig:leoptics}), it is possible to continuously cover the FoV with four units of nine-segment MuLE.
Accordingly, we can achieve an FoV of $\sim$1~sr with only four imagers (with the total imager area of 10 $\times$ 10~cm$^2$).

\section{Ray tracing simulation}

To evaluate the performance of the MuLE optics, including the surface roughness of the mirrors, misalignment of the LE mirror cells, and a realistic detector configuration, we performed a ray tracing simulation by modifying a previously built simulator \cite{Mitsuishi:2016uh}.
The values of the surface roughness and the mirror cell misalignment were taken from our past mirror fabrication \cite{Ezoe:2018hr}.
In this study, we treated only one parameter set because we aimed to evaluate the working principle of the MuLE optics.
The optimization of the parameters will be discussed in our next publication.

\subsection{Simulation setup}

The simulation was performed with the nine-segment MuLE configuration shown in Fig.~\ref{fig:le_module_config}.
One LE segment has a size of 10 $\times$ 10~cm$^2$.
Given that the support structures of the LE segments are necessary in a realistic design, a 0.5-cm margin was added around each LE segment.
Thus, the geometrical area of each LE segment becomes 9 $\times$ 9~cm$^2$.
The nine LE segments are tiled on a spherical surface with a radius $R=60$~cm.
The azimuthal rotation angle of each LE segment is incremented by 10$^{\bf\circ}$ from 0$^{\bf\circ}$ (central one) to 80$^{\bf\circ}$, as shown in Fig.~\ref{fig:le_module_config}.

Recently, some X-ray mirrors have been produced with a silicon--microelectromechanical systems (Si--MEMS) technology \cite{Ezoe:2018hr} that is a precise and a less expensive technique applicable to the LE optics.
We assumed the use of the Si--MEMS technology in the MuLE optics and chose parameters that could be used for manufacturing with current technologies.
The thickness of the silicon wafer was 300~$\mu$m, and the pore size was 20 $\times$ 20~$\mu$m$^2$.
Since the spacing between adjacent pores was 40~$\mu$m, the open fraction of the aperture was 25\%.
To keep the structural strength of the Si--MEMS mirror, radial spokes with widths of 300~$\mu$m were added every 15$^{\rm \circ}$. 
This reduced the aperture ratio to 21\%.
Compared with the standard LE mirror made of glass material, the thickness is about one-third, but the other properties such as the point spread function are comparable.

In the simulations, X-rays originating from the nine LE segments were captured by a 4~k $\times$ 4~k pixel imager with a sensitive area of 6.144 $\times$ 6.144~cm$^2$ (i.e., with a 15-$\mu$m pixel size) centered at the focal point of $f=30$~cm.
The state-of-art complementary metal-oxide semiconductor (CMOS) technology allows us to use low-noise pixel imagers without cooling.
GPixel's CMOS sensors represent these types of devices \cite{Wang:2018hy}.
The 15-$\mu$m pixel size corresponds to the arc length of 10 arcsec in the sky coordinate system.
It is small enough compared with the imaging quality of the Si--MEMS LE optics.
The detailed values of parameters for the ray tracing simulation are summarized in Table~\ref{tab:sim_params}. 

\begin{threeparttable}
\begin{center}
\caption{Parameters of the ray tracing simulation
\label{tab:sim_params}
}
\begin{tabular}{ll}
\hline
Parameter&Value\\
\hline
Scan energy ($E$) & 0.5 to 3.5~keV (0.5~keV step)\\
Scan angle of photons ($\Theta_x$, $\Theta_y$) \tnote{*} & 0 to 10$^{\bf\circ}$ (2$^{\bf\circ}$ step)\\
Thickness of lobster-eye (LE) mirror ($\ell$) & 300~$\mu$m\\
Radius of LE sphere ($R$) & 60~cm\\
Focal length of LE optics ($f$) & 30~cm\\
Open fraction of cells  ($\eta$) \tnote{$\dagger$} & 0.21\\
Cell size ($w\times w$) & 20 $\times$ 20~$\mu$m$^2$\\
LE mirror segment size ($L\times L$) & 10 $\times$ 10~cm$^2$\\
LE mirror effective area ($L_\mathrm{e} \times L_\mathrm{e}$) & 9 $\times$ 9~cm$^2$\\
Mirror coating material & Pt\\
Mirror surface roughness \tnote{$\ddagger$} & 1~nm (rms)\\
Mirror point spread function \tnote{$\ddagger$} & 10~arcmin (FWHM) \\
Imager size \tnote{\S} & 6.144 $\times$ 6.144~cm$^2$\\
\hline
\end{tabular}
\begin{tablenotes}
\item[*] Scan angles are measured from the center of the field-of-view of each LE segment.
\item[$\dagger$] Shadows induced by the radial spokes are included.
\item[$\ddagger$] These values were obtained from our Si--MEMS manufacturing experience.
\item[\S] 4 k $\times$ 4 k square pixels (pixel size = 15~$\mu$m).
\end{tablenotes}
\end{center}
\end{threeparttable}

\subsection{Image response of a point source}

\begin{figure}
\begin{center}
\includegraphics*[width=13cm]{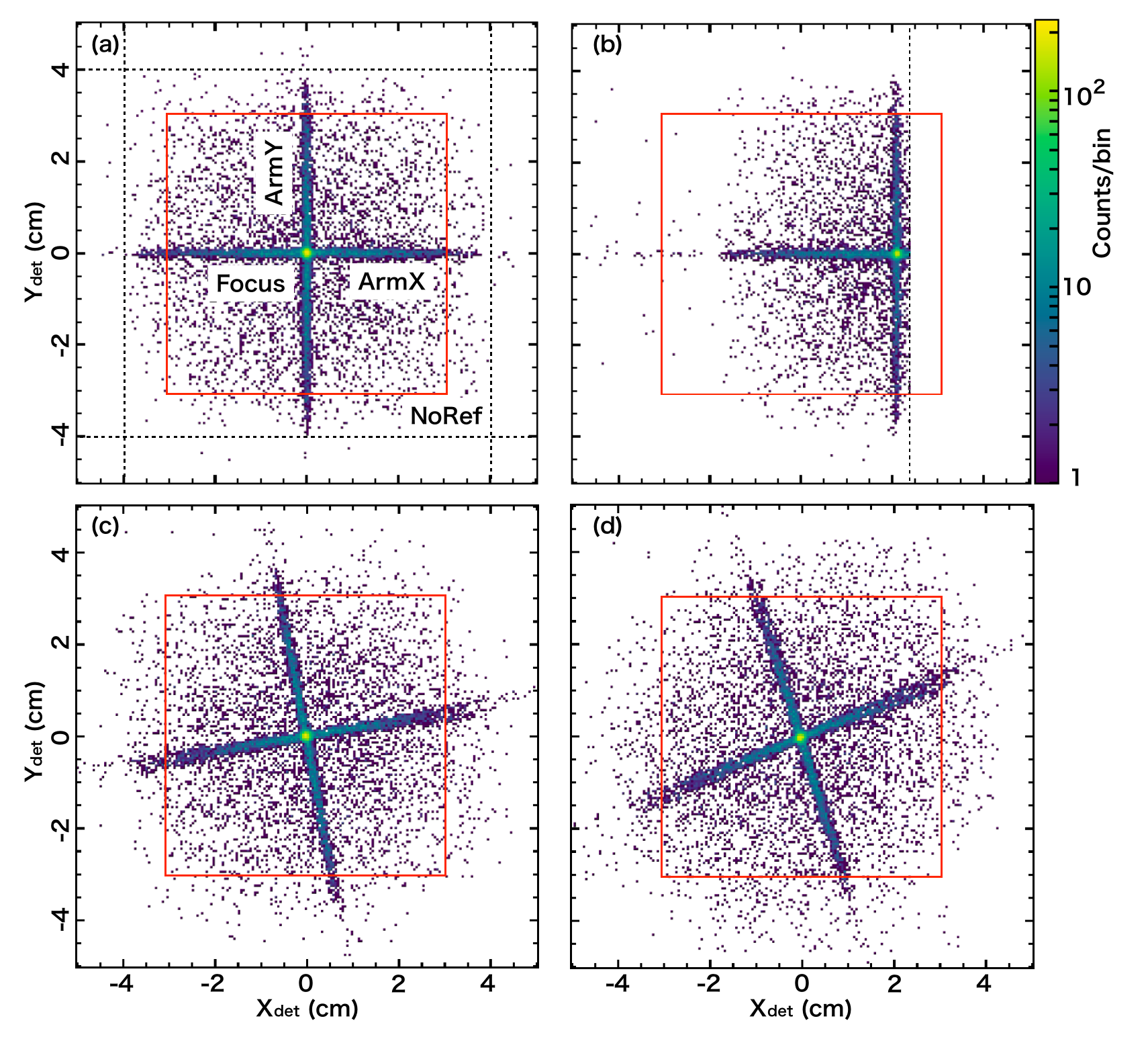}
\end{center}
\caption{(a) Simulated image of a point source focused by ID00-$\phi$00 with an incident photon angle of $\Theta_x=\Theta_y=0^{\rm\circ}$.
See text for detail on "Focus" and "ArmX/Y". 
The dashed lines show the boundary beyond which there are not non-reflected (NoRef) photons. 
The red rectangle shows the size of the imager. 
(b) ID00-$\phi$00 with $\Theta_x=4^{\rm\circ}$ and $\Theta_y=0^{\rm\circ}$.
The dashed line shows the image boundary produced by the edge of the LE segment.
(c) ID20-$\phi$10 with $\Theta_x=\Theta_y=0^{\rm\circ}$. 
(d) ID22-$\phi$20 with $\Theta_x=\Theta_y=0^{\rm\circ}$.
The color bar shows the counts-per-bin on a logarithmic scale.
All of the images were reconstructed with $\sim$12,000 photons at 0.6~keV.
\label{fig:pointsource_images}
}
\end{figure}

First, we simulated the image response of a point source focused by the ID00-$\phi$00 LE segment.
Figure~\ref{fig:pointsource_images}a shows the image response of a point source with an incident angle of $\Theta_x=\Theta_y=0^{\rm\circ}$ with respect to the central optical axis of the ID00-$\phi$00 segment.
Approximately 15\% of the detected photons at 0.6~keV were scattered twice on the adjacent walls of the LE cells and focused at the center of the imager (marked as "Focus" in the figure).
Approximately 48\% of the photons were scattered once on the cells and concentrated in the cross-like arm foci (these are marked as "ArmX" and "ArmY").
The remaining 37\% of the photons were dropped through the cells directly to the imager (these are marked as "NoRef").
The boundary limit angle beyond which NoRef photons do not exist is defined by $\theta_\mathrm{lim}=\tan^{-1}(w/\ell)=3.81^{\rm\circ}$, and corresponds to 4~cm ($=R\sin\theta_\mathrm{lim}$) on the imager.

Second, we considered the point source with an incident angle of $\Theta_x=4^{\rm\circ}$ and $\Theta_y=0^{\rm\circ}$ with respect to the central optical axis of ID00.
Figure~\ref{fig:pointsource_images}b shows the image of the source clearly shifted to the right compared with Fig.~\ref{fig:pointsource_images}a.
Only half of the image was detected in the X-axis direction, but it was sufficiently detected even at the edge of the FoV.
In realistic configurations used in X-ray astronomy, the missing half of the X-ray images could be detected by another MuLE unit given that the FoVs are tiled withoug gaps, as shown in Fig.~\ref{fig:fieldofview}, i.e., the reduction of the effective area can be almost mitigated.
The boundary created by the edge of the LE segment is clearly seen in  Fig.~\ref{fig:pointsource_images}b at $X_{\rm det}=2.4~{\rm cm}$ for $\Theta_x=4^{\rm\circ}$.
See Appendix~\ref{app:lobster_eye_optics} for a detailed description of the edge of the LE segment.

Finally, we simulated the point source images focused by the ID20 and ID22 segments.
Figure~\ref{fig:pointsource_images}c shows the image focused by the ID20-$\phi$10 segment.
The image response was similar to that of ID00 but was rotated 10$^{\rm\circ}$ as the LE segment rotated.
Figure~\ref{fig:pointsource_images}d shows the image focused by the ID22-$\phi$20 segment.
The cross-like images in both the ID20 and ID22 segments were clearly seen.
This implied that the images from any LE segment could be detected.

As expected, defocus aberration was observed at the edge of the CMOS image sensor for the ID20 and ID22 segments, given that the focal plane was tilted in these segments.
The worst case of the defocus aberration appeared at the diagonal edge of the CMOS imager for the ID22 segment.
At that point, the focal length was $\sim$1.1~cm shorter than that for the true focal length $f=30~{\rm cm}.$
The defocus corresponds to $\pm$8.0~arcmin aberration in the sky coordinate system.
Since this is almost comparable to the FWHM size (10~arcmin) of a point source focused by the Si--MEMS mirrors, the defocus was not a problem in our configuration.

\subsection{Effective area}

\begin{figure}
\begin{center}
\includegraphics*[width=13cm]{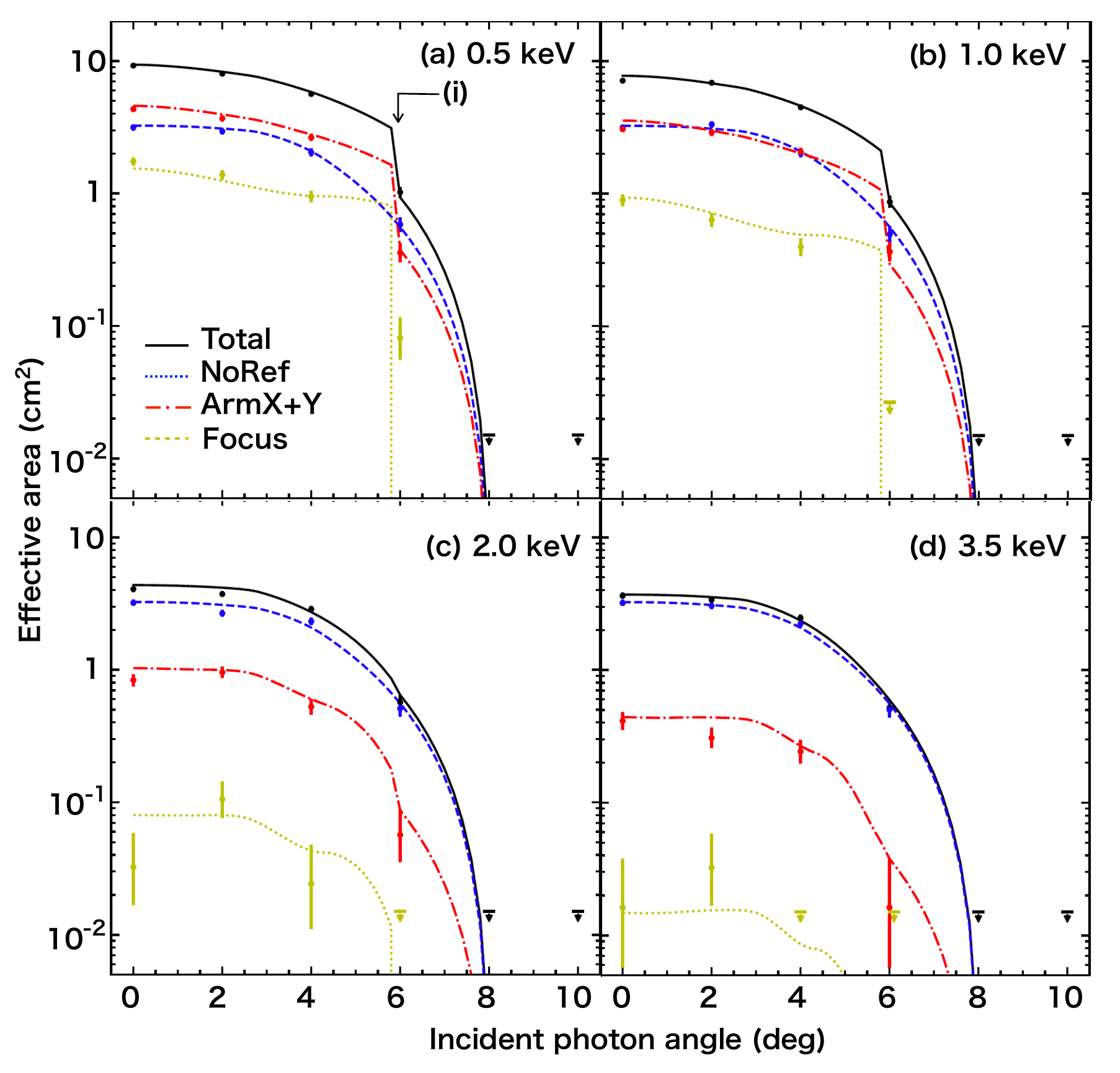}
\end{center}
\caption{Effective area as a function of the incident photon angle $\Theta_x$ with $\Theta_y=0^{\rm \circ}$ for ID00 at (a) 0.5, (b) 1.0, (c) 2.0, and (d) 3.5~keV. 
Only the photons collected by the CMOS imager are taken into account.
The data points show the results of the ray tracing simulation, and the curves show the analytic calculation.
Point (i) indicates the edge angle where the Focus is shfted off the edge of the image sensor.
\label{fig:effective_area_angle}
}
\end{figure}

The mirror effective areas were also derived from the ray tracing simulation.
Figures~\ref{fig:effective_area_angle}a--\ref{fig:effective_area_angle}d show the effective areas of ID00 as a function of the incident photon angle measured from the optical axis of the LE segment for 0.5, 1.0, 2.0, and 3.5~keV, respectively.
In this calculation, the size of the CMOS sensitive area was taken into account, but the quantum efficiency of the imager was not since the efficiency is almost 100\% in this energy band.
The simulations were conducted based on discrete calculations within the angle range of $\Theta_x$ at 2$^{\rm\circ}$ steps.
The reason for including NoRef in the figures of the effective areas was that the LE optics had two functions: a focusing mirror (ArmX/Y and Focus) and a collimator (NoRef).
Since the density of X-ray objects in the sky is sparse, if no other object is in the FoV, NoRef is identified as X-rays from the target object.

\begin{figure}
\begin{center}
\includegraphics*[width=13cm]{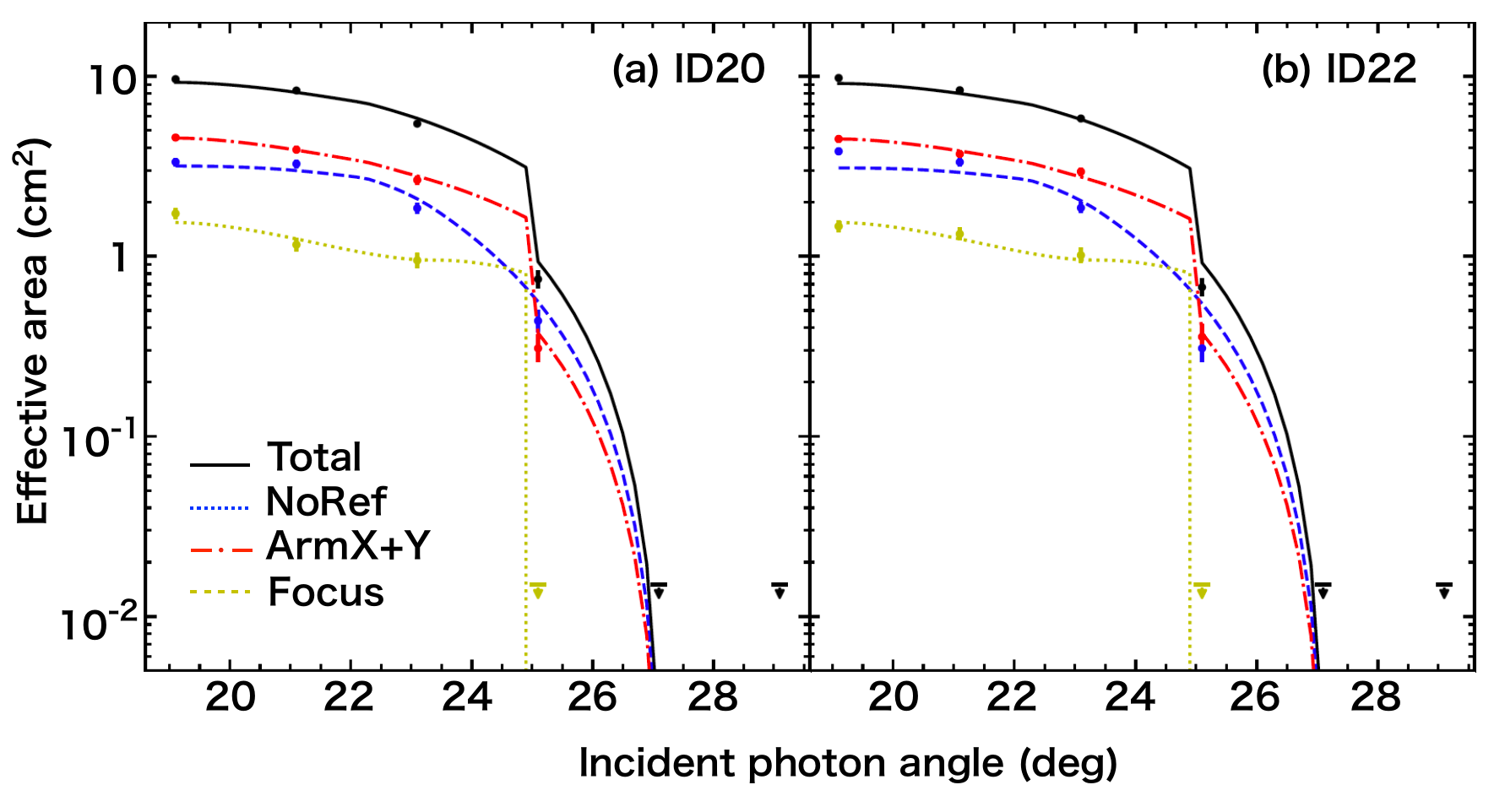}
\end{center}
\caption{Effective area as a function of the incident photon angle $\Theta_x$ for (a) ID20 with $\Theta_y=0^{\rm\circ}$ and (b) ID22 with $\Theta_y=19.1^{\rm\circ}$ at 0.5~keV. The angles are measured from the optical axis of ID00. The curves of the effective areas are symmetrical about 19.1$^{\rm\circ}$.
\label{fig:id2022_area_angle}
}
\end{figure}

The curved lines shown in Figs.~\ref{fig:effective_area_angle} were analytically calculated effective area in combination with the mirror reflectivity \cite{Henke:1993gz}.
The detailed procedure of the analytic calculation is summarized in Appendix~\ref{sect:analytic_estimation_of_effective_area}.
The discontinuity marked (i) in Fig.~\ref{fig:effective_area_angle}a shows the angle where the Focus is shifted off the edge of the CMOS.
The effective area for ID20 and ID22 at 0.5~keV are shown in Figs.~\ref{fig:id2022_area_angle}a and \ref{fig:id2022_area_angle}b respectively.
Given that the difference between ID20/22 and ID00 is originated only in the tilt angle of the X-ray images, the curves of the effective area look very similar to each other.
To clarify the characteristics of the nine-segment MuLE optics only, the vignetting is shown in Appendix~\ref{sect:vignetting_of_mule}. 

\begin{figure}
\begin{center}
\includegraphics*[width=10cm]{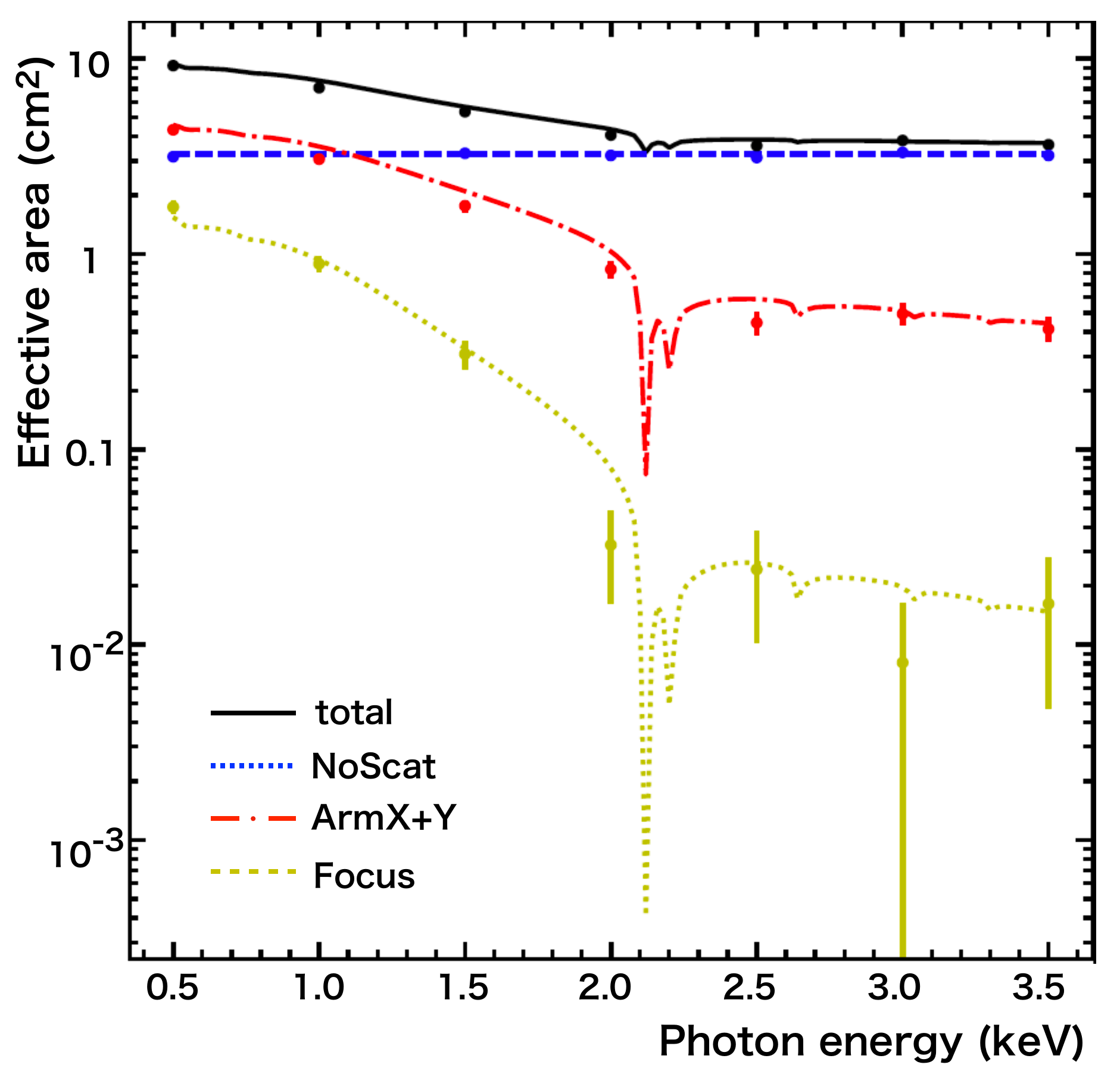}
\end{center}
\caption{Effective area as a function of photon energy for ID00 with an incident angle of $\Theta_x=\Theta_y=0^{\rm \circ}$.
\label{fig:effective_area_energy}
}
\end{figure}

The effective areas of the mirror as a function of the incident photon energy for the ID00 segment with an incident angle of $\Theta_x=\Theta_y=0^{\rm\circ}$ are shown in Fig.~\ref{fig:effective_area_energy}.
The ray tracing simulation was performed for different energies at every 0.5~keV from 0.5 to 3.5~keV.
The curves of the effective areas derived from the analytic calculation are also shown in the figure.
While the effective area of NoRef was flat, the effective areas of ArmX/Y and Focus dropped rapidly as the energy increased.
The effective area of ArmX+Y was somewhat larger than that of NoRef below 1~keV.

\subsection{Source detection limit}

\subsubsection{Photon and background limit cases}

\begin{figure}
\begin{center}
\includegraphics*[width=13cm]{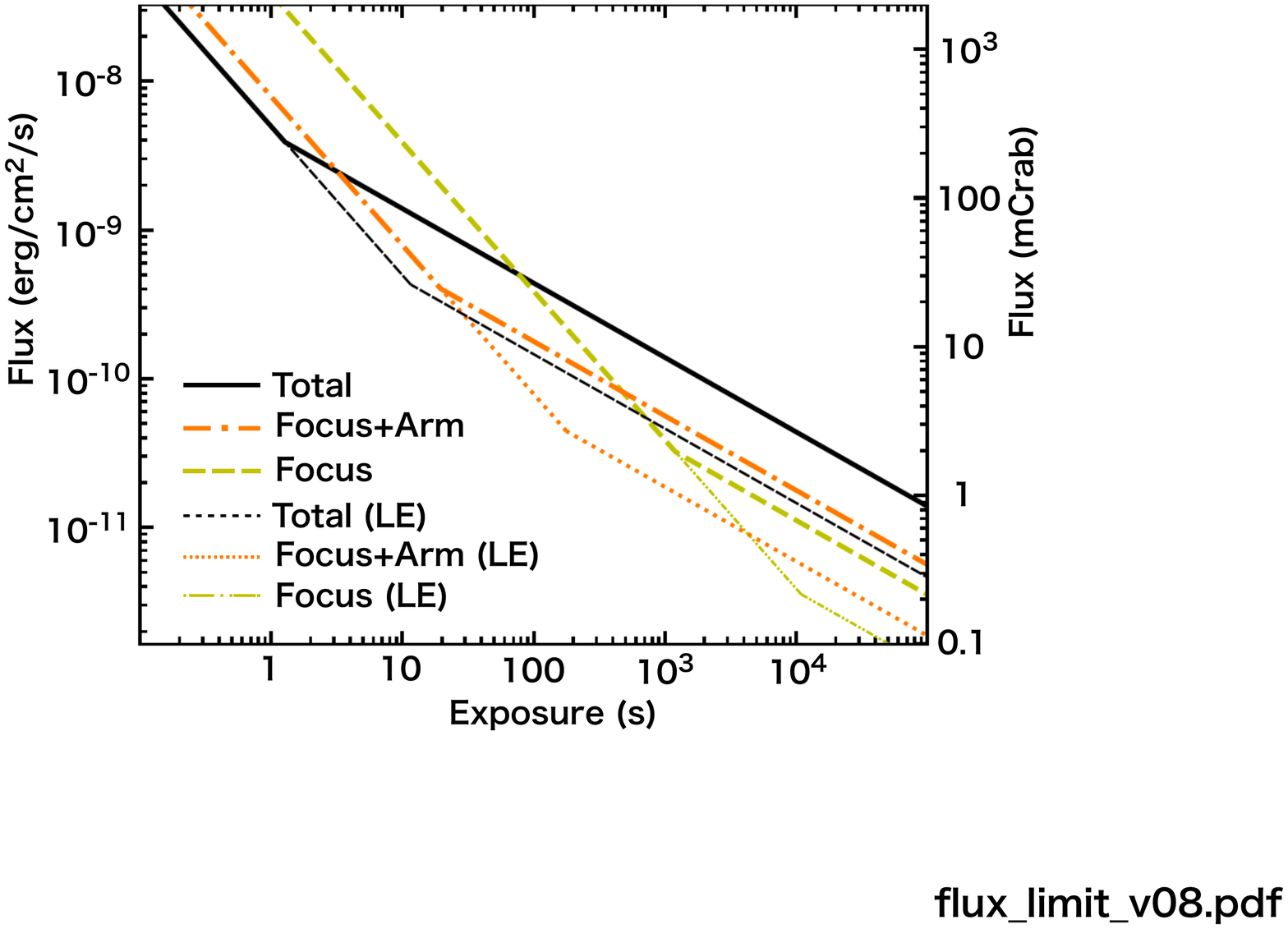}
\end{center}
\caption{The 5$\sigma$ detection limit of a point source at the center of the ID00 FoV in the 0.5 to 2.0~keV bandpass for the nine-segment MuLE optics. The 5$\sigma$ detection limit for the standard LE configuration is overlaid.
\label{fig:flux_limit}
}
\end{figure}

The source detection limit was determined by the signal-to-noise ratio of the X-ray photons on the imager.
In the MuLE optics, the most dominant noise is the diffuse CXB.
Fig.~\ref{fig:flux_limit} shows the 5$\sigma$ detection limits for Focus, Focus+ArmX/Y, and total (Focus+ArmX/Y+NoRef) when a point source was located at the center of the ID00 FoV.
To extract foreground and background photons in the region of ArmX+Y and Focus, we selected the photons in the strip regions along the arm foci (widths of 0.2~cm).
The strip width was not optimized but was adequately large enough to collect the photons focused by the LE segments even in the cases in which the image suffered defocus.
Throughout this study, we assumed a Crab-like spectrum \cite{Kirsch:wl} for a point source characterized by a power-law photon index of 2.07, normalization of 8.26 photons keV$^{-1}$ cm$^{-2}$ s$^{-1}$ at 1~keV, and an absorption of $N_H=4.5 \times 10^{21}$ cm$^{-2}$.

The flux limit was governed by the number of photons for shorter exposures (photon limit), and was proportional to $t^{-1}$, where \textit{t} is the exposure time.
Conversely, the flux limit was governed by the CXB photons in the cases of longer exposures (background limit), and was proportional to $t^{-0.5}$ because the number of background photons obeyed Poisson's Law.

Figure~\ref{fig:flux_limit} also shows the 5$\sigma$ detection limits for the standard LE configuration in which the size and properties of the LE segments were exactly the same but the images were not multiplexed.
Mathematically, the amount of background was reduced to one-ninth from that of the MuLE configuration.
The difference between the two configurations only appears in the background limit case as shown in Fig.~\ref{fig:flux_limit}.

\subsubsection{Confusion case of background point source}

Another possible weak point relevant to the nine-segment MuLE configuration is its large FoV which causes contamination of bright background sources in the imager.
The detection ability of a faint source is easily affected by a bright background source located in any of the nine FoVs.
We considered a background point source which was 0.5$^{\rm\circ}$ away from the object, which we observed at the center of FoV to evaluate its effects.
We calculated the detection limit change owing to the bright object for the case of Focus + ArmX/Y described in Fig.~\ref{fig:flux_limit}.

\begin{figure}
\begin{center}
\includegraphics*[width=12cm]{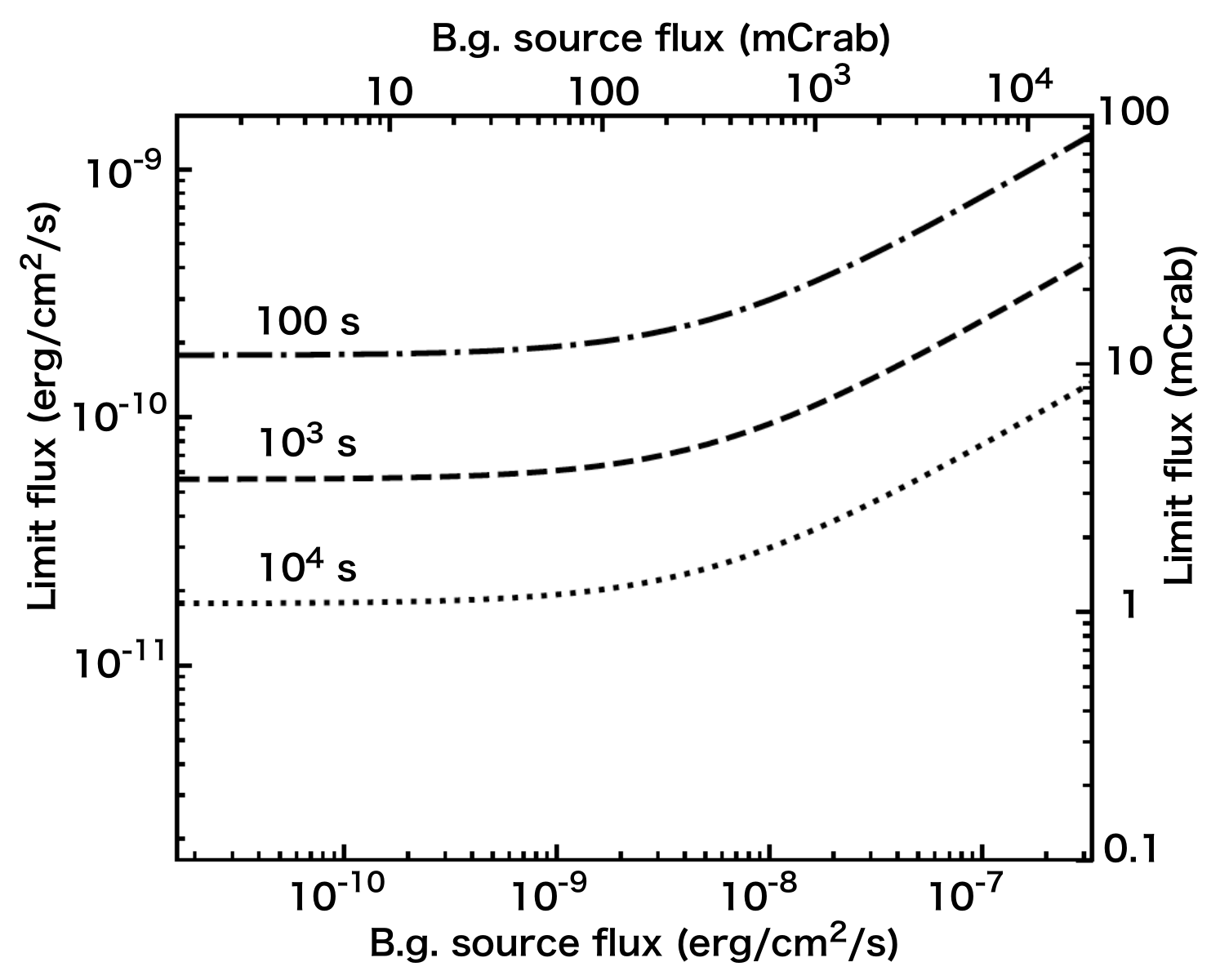}
\end{center}
\caption{Degradation of the 5$\sigma$ detection limit in the 0.5 to 2~keV band owing to the background (b.g.) source of the 0.5$^{\rm\circ}$ separation distance in the case of the nine-segment MuLE optics.
\label{fig:confusion}}
\end{figure}

Figure~\ref{fig:confusion} shows the detection limits for the observation times of 100, 10$^3$, and 10$^4$~s.
When the brightness of the background point source was brighter than 100~mCrab, the detection limit was degraded.
This is because the flux limit was governed by CXB, which is almost equivalent to a 100~mCrab source.

There are $\sim$30 objects in the entire sky that are brighter than 100 mCrab in the X-ray band.
For the nine-segment MuLE configuration that we considered, the FoV was about 666 deg$^2$ (nine $8.6^{\rm\circ}\times8.6^{\rm\circ}$ FoVs), which corresponds to 1.6\% of the entire sky and contains $\sim$0.5 bright objects on average.
Since many bright X-ray objects are distributed along the galactic plane, they are not a fatal background when we observe the region of the high galactic latitude.

\subsection{FoV discrimination based on image response}

To evaluate the power of the FoV discrimination by the cross-like image response, we also employed the ray tracing simulation.
This problem is converged to a problem that pertained to the determination of the azimuthal rotation angle of the cross-like image.

\subsubsection{Method used to determine the azimuthal rotation angle of cross-like image}
\label{subsubsect:method_to_determine_the_phase_angle}

We considered the nine-segment MuLE configuration shown in Fig.~\ref{fig:le_module_config}.
When the flux from a transient object exceeds the detection limit, at least one image is captured.
At this moment, it is unclear which LE segment (ID00-$\phi$00 to ID2$\overline{\rm2}$-$\phi$80) focused the image.
In consideration of all possibilities, the image is subjected to nine different operations to identify the LE segment that was involved.
The procedure that we employed is as follows.

\begin{figure}
\begin{center}
\includegraphics*[width=10cm]{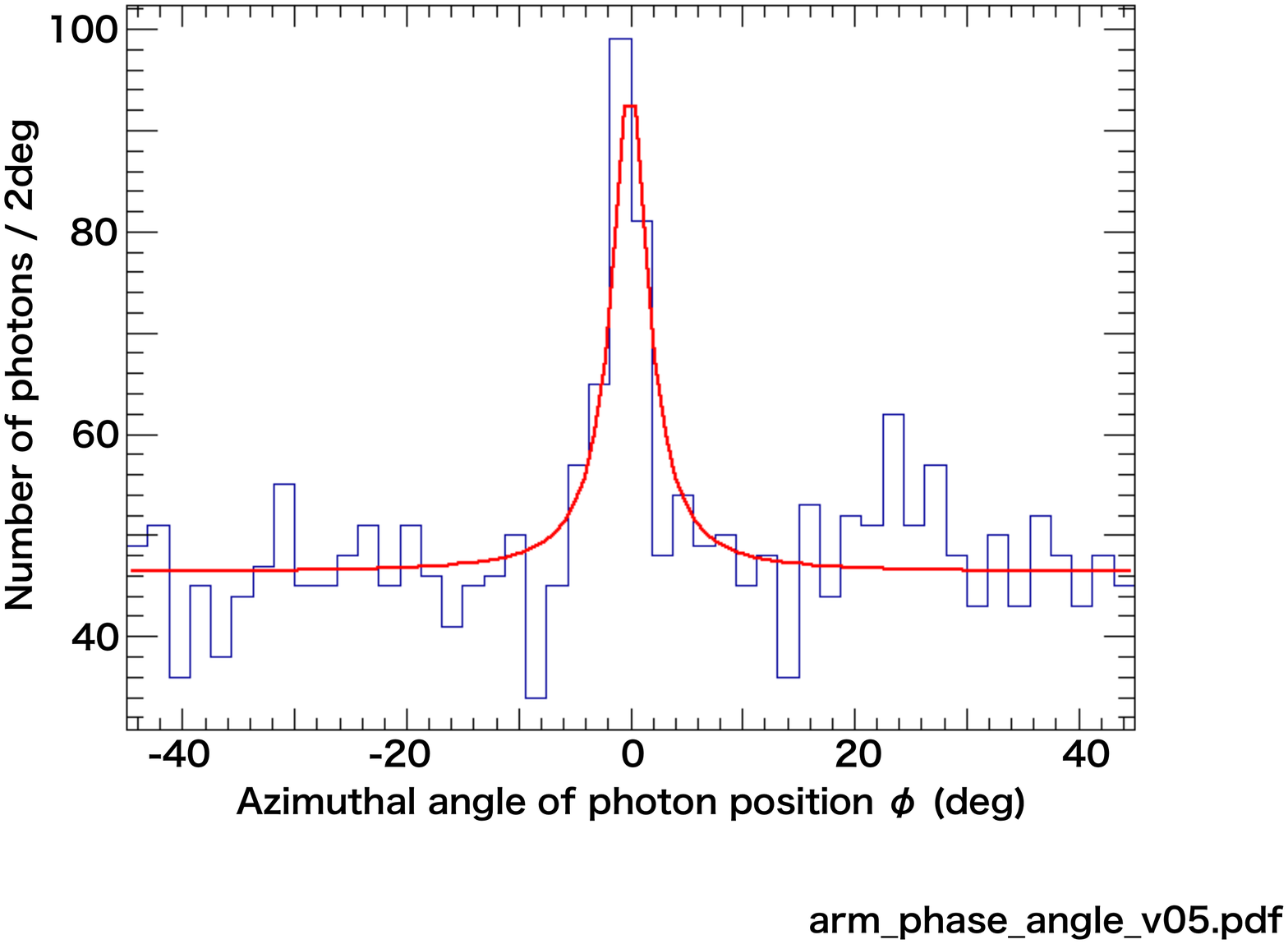}
\end{center}
\caption{
(Histogram) Typical distribution of the azimuthal angles of photon positions for the 250 source and 100~s CXB photons for the nine-segment MuLE optics.
(Curve) The best fit result of the plotted distribution with the template response.
}
\label{fig:arm_phase_angle}
\end{figure}

\begin{enumerate}
\renewcommand{\labelenumi}{\arabic{enumi}).}
\item
    For LE segments other than ID00, image distortion should be corrected first given that the imager was tilted with respect to the tangential plane at the center of the LE segment.
    The distortion correction produced eight different images.
    Details of the correction are described in Appendix~\ref{sect:correction_method_of_images}. 
    Currently, there are a total of nine images.

\item
    By identifying the center of gravity of the entire photons, the position of the transient source on the imager O$_i$ ($i=1, ..., 9$) is determined in all nine images.
    
\item
    The position of each photon P$_{i,m}$ ($m=1,..., N$) is recorded, where $N$ is the number of total photons. Then, the azimuthal rotation angles $\phi_{i,m}$ of the vector from O$_i$ to P$_{i,m}$ are calculated. The azimuthal rotation angles $\phi_{i,m}$ are measured from the azimuthal rotation angle $\phi_i$ of the square cells of the LE segment.

\item
    The azimuthal rotation angles $\phi_{i,m}$ are filled in a histogram between $-45$ and $+45^{\rm\circ}$ given that the cross-like point source image has four-fold rotational symmetry.
    Only the photons in a ring region of the radius between 0.15 and 3~cm are sampled concentrically around O$_i$.
    Figure~\ref{fig:arm_phase_angle} shows an example of the histogram for the case of ID00-$\phi$00.
    Herein, there are a total of nine histograms.

\item
    The point source responses prepared in advance for all nine LE segments are fitted to a histogram, and the goodness-of-fit was found based on the maximum likelihood estimation.
    The point source response is generated by the ray tracing simulation with sufficient statistics for more than 100,000 photons: CXB photons are not included.
    The response was modeled with a Lorentzian function and a constant as according to
    \begin{equation}
        f(x) = S \frac{\Gamma/2}{(\phi-\phi_i)^2 + (\Gamma/2)^2} + N.
    \end{equation}
    The parameter $\phi_i$ was fixed to the azimuthal rotation angle of the LE segment cells, and the half-width was fixed to the value $\Gamma/2=1.975^{\rm\circ}$ derived from the simulation.
    The other two parameters, Lorentzian normalization $S$ and the constant value $N$, were free in the fit.
    An example of the fit is shown in Fig.~\ref{fig:arm_phase_angle}.

\item
    The operations are performed for all the nine images, and the one with the highest $S/N$ is selected as the LE segment from which the point source originated.
\end{enumerate}

\subsubsection{FoV determination for ID00 and ID22}

\begin{figure}
\begin{center}
\includegraphics*[width=10cm]{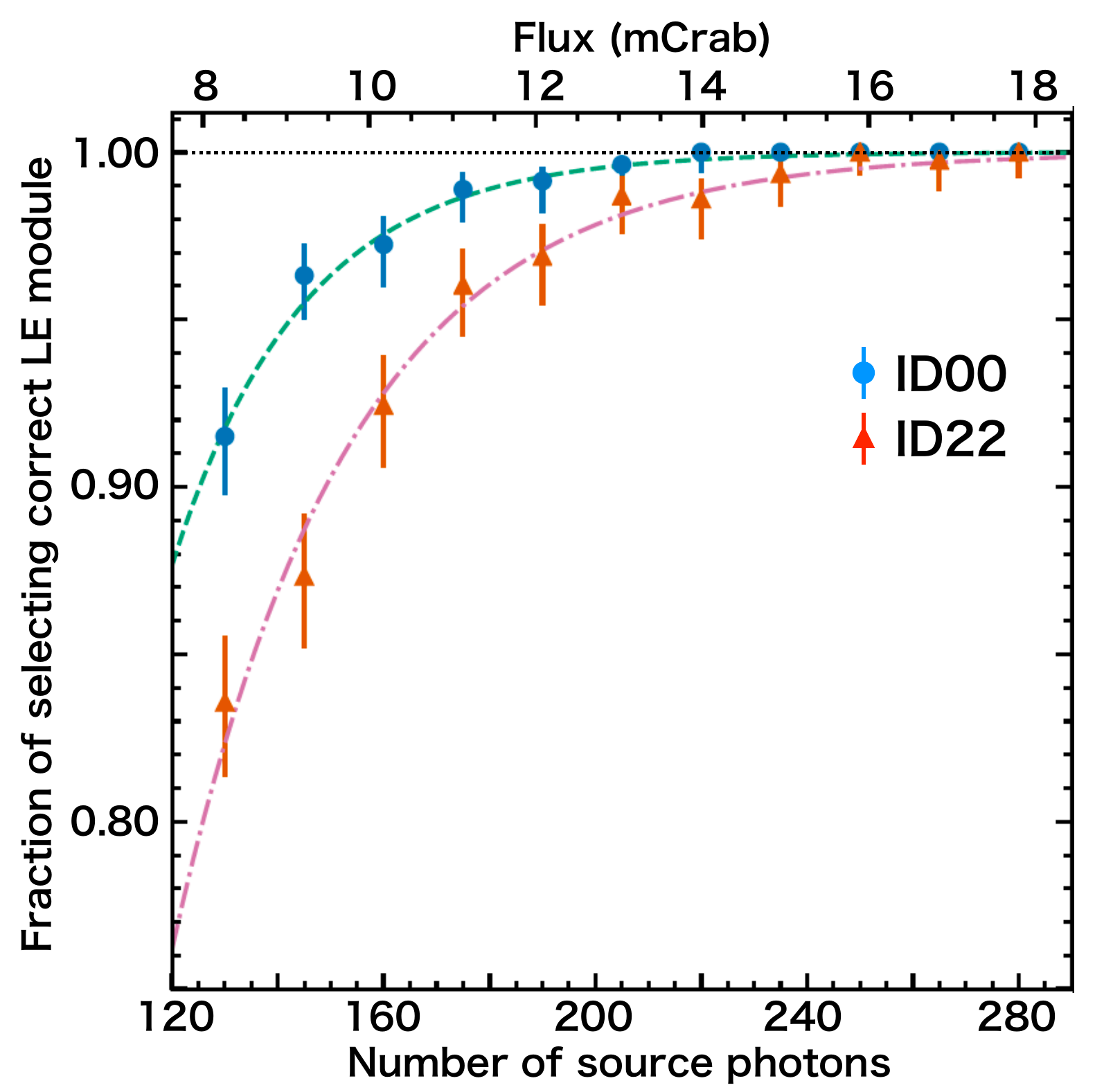}
\end{center}
\caption{
Fraction associated with the selection of the correct LE segment for 0.6~keV X-rays detected with the nine-segment MuLE optics.
}
\label{fig:fov_determination}
\end{figure}

We performed ray tracing simulations for a transient with a duration of 100~s to evaluate if we could localize its position as a function of the source flux.
In this study, the number of CXB photons was fixed for 100~s observation, but the number of X-ray photons from the transient source was varied.
Using the method described in \S\ref{subsubsect:method_to_determine_the_phase_angle}, the segment that focused the photons on to the imager was determined from the simulated data.
Figure~\ref{fig:fov_determination} shows the fraction associated with the selection of the correct LE segment as a function of the number of source photons.
Each data point was the average of 350 to 850 trials. Corresponding error bars are also plotted.
For simplicity, this study was conducted with 0.6~keV photons.

The number of photons required to achieve 95\%, 99\%, and 99.7\% correct outcome rates for ID00 were 142, 182, and 212, respectively.
In combination with Fig.~\ref{fig:flux_limit}, the position of a point source was correctly determined in 97\% of the events at the 5$\sigma$ detection threshold for 100~s observations.
Even if the correct LE segment could not be determined from the data, the source position on the imager was determined with an FWHM accuracy of $\sim$10~arcmin and can be narrowed down to nine points in the sky coordinate system.
Furthermore, if the correct LE segment can be identified with a deep learning approach, the determination accuracy may be improved.

Similarly, the number of photons required to achieve 95\%, 99\%, and 99.7\% of correct outcome rates for ID22 were 172, 226, and 266, respectively.
The reason for which the correct outcome rate being lower than that of ID00 at the same photon numbers is that the image sensor of ID22 was tilted with respect to the tangential surface of the LE segment and the arm of the cross-like response was blurred owing to the defocus effect.
It would be useful to perform a more detailed survey to assess the performance of the MuLE optics. However, this is beyond the scope of this study and will be described in our next publication.

\section{Discussions and outlook}

This study described the working principle of the MuLE optics in which multiple LE segments focused X-rays onto a single imager. 
This configuration reduced the number of image sensors considerably and thus overcame a disadvantage of the LE optics.
A ray tracing simulation was performed to evaluate the properties of the MuLE optics based on the assumption of a nine-segment configuration.
In the simulation, only the existing technologies (Si--MEMS mirrors and a CMOS image sensor) that will help with the construction of an inexpensive and accurate enough wide-field X-ray monitor in the near future were assumed.

When the focal length of 30~cm and an area spanning 9 $\times$ 9~cm$^2$ of an LE segment were used, the total effective area at 1~keV was calculated to be 8~cm$^2$ at the center of the FoV, and about 4~cm$^2$ at the edge of the FoV ($\Theta=\pm4.3^{\rm\circ}$).
The 5$\sigma$ detection limit in the 0.5- to 2-keV band for a transient with a duration of 100~s at the center of FoV was  $\sim$2 $\times$ 10$^{-10}$ erg cm$^{-2}$ s$^{-1}$ (10~mCrab).
The ability to determine the correct position achieved a 99.7\% level for a 14 to 17~mCrab point source with a duration of 100~s.
Thus, we finally conclude that the MuLE optics can be used to implement a wide FoV transient monitor with sufficient sensitivity.

Given that the MuLE configuration is an easiest way to reduce considerably the number of image sensors, it is considered to be effective for a small satellite with limited resources or a small observatory on-board the International Space Station.
With the use of the three units of the nine-segment MuLE with $f=30$~cm, as presented in this study, it is possible to cover a 0.75~sr of an FoV with a microsatellite with a volume of 50 $\times$ 50 $\times$ 50~cm$^3$.
With 16 satellite sets, the entire sky can be covered.
Using lightweight and inexpensive Si--MEMS technology and by reducing the number of imaging devices with MuLE, the price per MuLE unit can be reduced considerably. Accordingly, the establishment of a constellation of these types of microsatellites is possible.

The ability to cover the entire sky at all times with the satellite constellation will have a major impact in the multimessenger and taime-domain astronomy.
If the focal length is reduced by half to 15~cm, the number of satellites in the constellation can be reduced to four, though the sensitivity will drop.
In addition, given that the MuLE configuration that we described in this study can achieve about 1~mCrab at 10$^4$~s, it can be used as an all-sky monitor, such as \textit{MAXI} or \textit{RXTE/ASM}.
Since the position is known in advance, for a known source, it is not necessary to identify the azimuthal rotation angle of the cross-like image, and the point source can be determined using only the location on the image sensor.
By optimizing the parameters, such as the increase of the thickness of the Si--MEMS mirror, we can fabricate more sensitive all-sky monitors.
In a future publication, we will discuss parameter optimization and examine the detailed performance of those configurations.

\appendix    

\section{Boundaries in the lobster-eye optics}
\label{app:lobster_eye_optics}

\begin{figure}
\begin{center}
\includegraphics*[width=15cm]{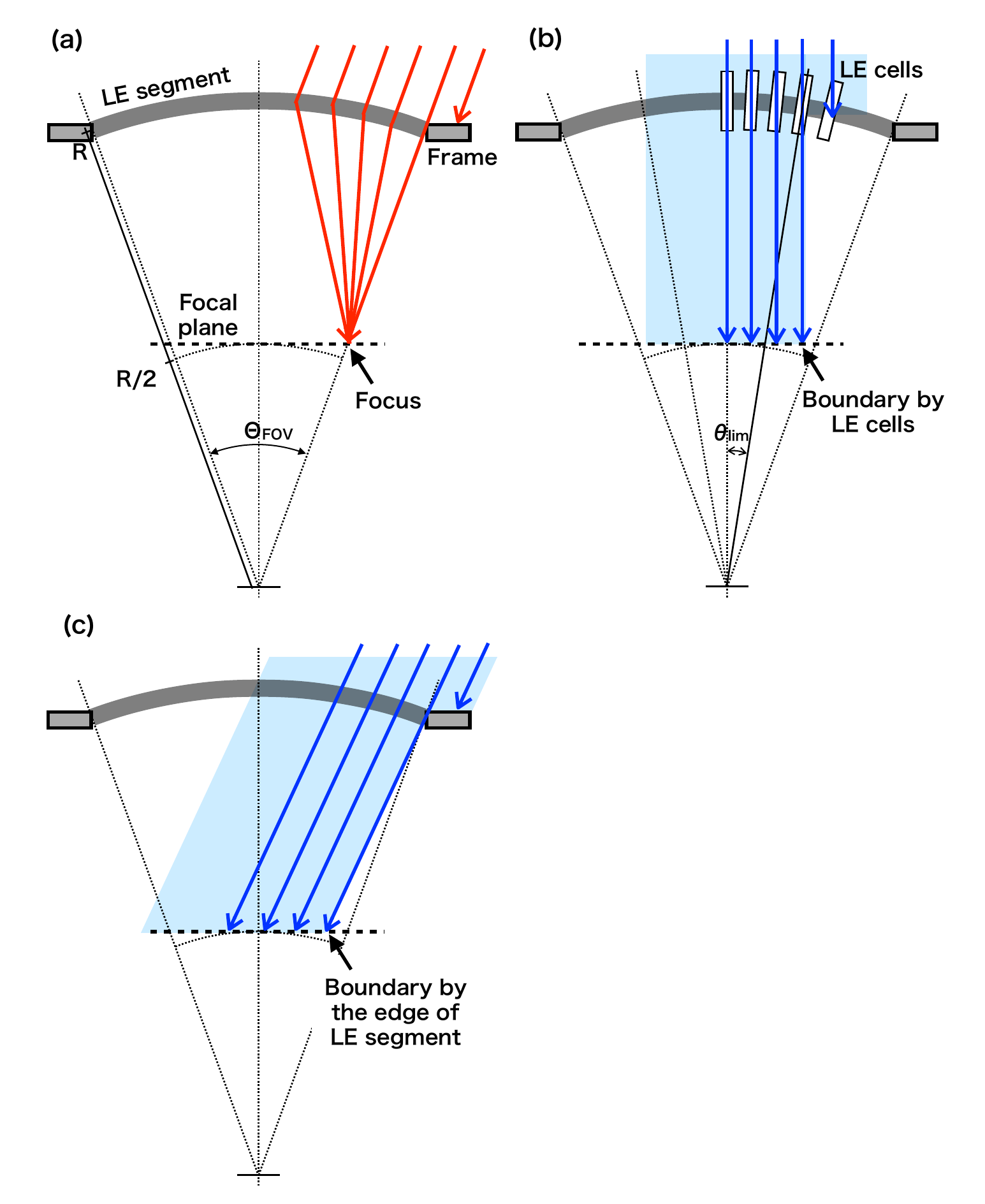}
\end{center}
\caption{
(a) The definition of FoV of the LE optics.
(b) The boundary produced by the LE cells and the limit angle $\theta_\mathrm{lim}$.
(c) The boundary produced by the edge of the LE segment.
These figures are exaggerated for readability.
The boundaries (b) and (c) are common for the NoRef photons and the unfocused direction of the Arm foci.
}
\label{fig:le_boundaries}
\end{figure}

Since the LE segments and LE hollow cells have a finite size, various boundaries appear in the LE optics.
Here, we explain the origins of some important boundaries.
For the specific numerical values shown in this section, the same parameters used in the simulation were used.
Figure~\ref{fig:le_boundaries}a shows the definition of the LE FoV.
It is defined that the center of the cross-like image is exactly on the line connecting the edge of the LE segment and the center of curvature.
With the parameters used in our simulation, FoV becomes $\Theta_{\rm FOV} \sim L_e/R \; \mathrm{rad} = 8.6^{\rm\circ}$.
Figure~\ref{fig:le_boundaries}b shows the boundary limited by the LE hollow cells for the photons that pass through without reflection.
This is the boundary visible in Fig.~\ref{fig:pointsource_images}a.
With our LE parameter, the limit angle becomes $\theta_\mathrm{lim}=\tan^{-1}(w/\ell)=3.81^{\rm\circ}$.
Figure~\ref{fig:le_boundaries}c shows the boundary limited by the support structure (frame) of the LE segment for the photons that pass through without reflection.
There are no non-reflected photons outside the boundary as seen in Fig.~\ref{fig:pointsource_images}b.
As observed from Fig.~\ref{fig:le_boundaries}c, the location of this boundary is a function of the incident photon angle.

\section{Analytic estimation of the effective area}
\label{sect:analytic_estimation_of_effective_area}

We summarize herein the methodology to calculate the effective areas.
When X-ray photons enter a cell of an LE segment, some of them go through the cell without reflection; the others are reflected by the wall of the cell once, twice, or more times.
These photons can be categorized by the number of reflections \cite{Angel:1979tg}.
The fraction of each category is a function of the tilt angle of a cell $\theta_j$, where $j$ denotes $x$ or $y$.
In the case for which there are no reflections, the fraction is
\begin{eqnarray}
  f_0^i(\theta_j) = \begin{cases}
    1-\frac{\ell}{w}\tan(\theta_j) & \theta_j\leq\tan^{-1}(\frac{w}{\ell})\\
    0 &
    \theta_j>\tan^{-1}(\frac{w}{\ell})\\
  \end{cases}
\end{eqnarray}
In the case for which there is a single reflection
\begin{eqnarray}
  f_1^j(\theta_j) = \begin{cases}
    \frac{\ell}{w}\tan(\theta_j) &        \theta_j\leq\tan^{-1}(\frac{w}{\ell})\\
    2-\frac{\ell}{w}\tan(\theta_j) & 
    \tan^{-1}(\frac{w}{\ell})<\theta_j\leq\tan^{-1}(\frac{2w}{\ell})\\
    0 & \theta_j>\tan^{-1}(\frac{2w}{\ell})
 \end{cases}
\end{eqnarray}
In our setup shown in Table~\ref{tab:sim_params}, the boundary angles are $\tan^{-1}(w/\ell)$=3.81$^{\rm\circ}$ and $\tan^{-1}(2w/\ell)$=7.59$^{\rm\circ}$.

Using the photon fraction sorted by the number of reflections, the effective areas are derived as follows:
\begin{eqnarray}
A_{\rm NoRef}(\Theta_x,\Theta_y) &=& \frac{A \eta}{N^x N^y}
  \int_{\theta_x^{\rm min}}^{\theta_x^{\rm max}}
  f_0^x d\theta_x
  \int_{\theta_y^{\rm min}}^{\theta_y^{\rm max}}
  f_0^y d\theta_y\\
 A_{\rm ArmX}(E,\Theta_x,\Theta_y) &=& \frac{A \eta}{N^x N^y}
   \int_{\theta_x^{\rm max}}^{\theta_x^{\rm min}}
   f_0^x d\theta_x
   \int_{\theta_y^{\rm max}}^{\theta_y^{\rm min}}
   \xi(E,\theta_y)
   f_1^y d\theta_y\\
 A_{\rm ArmY}(E,\Theta_x,\Theta_y) &=& \frac{A \eta}{N^x N^y}
   \int_{\theta_x^{\rm min}}^{\theta_x^{\rm max}}
   \xi(E,\theta_x) 
   f_1^x d\theta_x
   \int_{\theta_y^{\rm max}}^{\theta_y^{\rm min}}
   f_0^y d\theta_y\\
A_{\rm Focus}(E,\Theta_x,\Theta_y) &=& \frac{A \eta}{N^x N^y}
  \int_{\theta_x^{\rm max}}^{\theta_x^{\rm min}}
  \xi(E, \theta_x)
  f_1^x d\theta_x
  \int_{\theta_y^{\rm max}}^{\theta_y^{\rm min}}
  \xi(E, \theta_y)
  f_1^y d\theta_y
\end{eqnarray}
where $A$ is the geometrical area $L_\mathrm{e} \times L_\mathrm{e}$, $\eta$ is the open fraction of the pore, $N^j$ is the normalization factor $\int d\theta_j$, and $\xi(E,\theta_j)$ is the reflectivity of the platinum-coated LE mirror with a surface roughness of 1~nm that refers to the X-ray database of the Lawrence Berkeley National Laboratory \cite{Henke:1993gz}.
The limit angles $\theta_j^{\mathrm{max}}$ and $\theta_j^{\mathrm{min}}$ are restricted by the edge of the LE segment, including the radial spokes and the CMOS sensor.
The limit angles vary as the incident photon angles $\Theta_x$ and $\Theta_y$ vary because the viewing angle of the edge changes.

\section{Vignetting of the MuLE optics}
\label{sect:vignetting_of_mule}

\begin{figure}
\begin{center}
\includegraphics*[width=15cm]{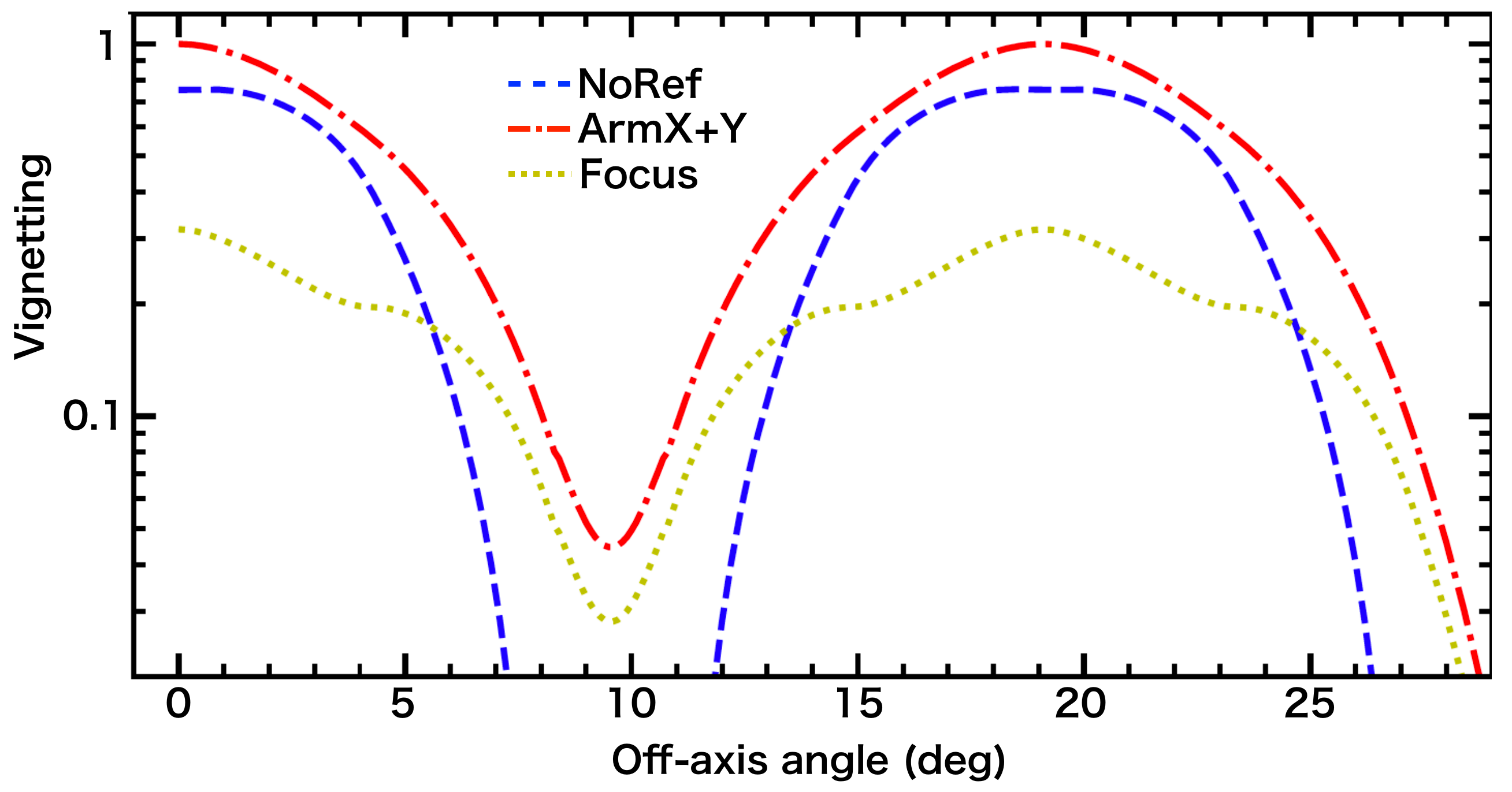}
\end{center}
\caption{Vignetting curve of the nine-segment MuLE optics as a function of off-axis angle along $\Theta_x$ with $\Theta_y=0$ for 0.5~keV photons. The vignetting is normalized by the value of ArmX+Y at 0$^{\rm\circ}$.
}
\label{fig:vignetting}
\end{figure}

We showed the effective areas of the MuLE configuration in Figs.~\ref{fig:effective_area_angle} and Figs.~\ref{fig:id2022_area_angle}, but they contain the effect by the finite size of the imaging detector.
It is worthwhile to show here the vignetting of the MuLE optics.
Figure~\ref{fig:vignetting} shows the vignetting curve along $\Theta_x$ with $\Theta_y=0^{\rm\circ}$ for 0.5~keV photons.
The second peak centered at 19.1$^{\rm\circ}$ was due to the ID20 segment.

\section{Correction method of elongated images detected with a tilted imager}
\label{sect:correction_method_of_images}

The images focused by any LE segment--except the ID00--are elongated because the focal plane imager is tilted with respect to the true focal plane of each LE segment.
To correct the elongated images, the following operation should be applied:
\begin{eqnarray}
\binom{x'}{y'} = A^{-1} \binom{x}{y},
\end{eqnarray}
where $(x, y)$ is the original position of a photon on an imager, and $(x',y')$ is the corrected position of the photon if the imager is located at the proper focal plane of the LE segment without the tilt angle.
The matrices \textit{A} are defined as
\begin{eqnarray}
A = \begin{pmatrix}k & 0 \\ 0 & 1\end{pmatrix},
  \begin{pmatrix}1 & 0 \\ 0 & k\end{pmatrix},
  \begin{pmatrix}\frac{k'+1}{2} & \frac{k'-1}{2} \\ \frac{k'-1}{2} & \frac{k'+1}{2} \end{pmatrix},\;
   {\rm and} \;
  \begin{pmatrix}\frac{k'+1}{2} & \frac{-k'+1}{2} \\ \frac{-k'+1}{2} & \frac{k'+1}{2} \end{pmatrix}
\end{eqnarray}
for ID20-$\phi$10/$\phi$50, ID02-$\phi$30/$\phi$70, ID22-$\phi$20/$\phi$60, and ID22-$\phi$40/$\phi$80, respectively, where $k=1/\cos{(\theta_t)}$ and $k'=1/\cos{(\theta_t')}$.
The tilted angles $\theta_t$ and $\theta_t'$ are defined as 
$4\tan^{-1}(L/2R)$ and $4\tan^{-1}(\sqrt{2}L/2R)$, respectively.

\acknowledgments 

This work was partially supported by the JSPS KAKENHI (Grant Number JP18K18775), Toray Science Foundation, and the budget for basic R\&D onboard equipment for future space science missions by the Advisory Committee for Space Science Japan.


\bibliography{mule}   
\bibliographystyle{spiejour}   

Toru Tamagawa is a chief scientist at RIKEN Center for Pioneering Research. He is the head of the high energy astrophysics laboratory. He received his PhD from the University of Tokyo in 2000. He has worked for the X-ray satellite missions in astrophysics: HETE-2, Suzaku, MAXI, Hitomi, GEMS/PRAXyS, IXPE, and XRISM.

\end{spacing}
\end{document}